\documentclass{article}[14pt]
\usepackage{amsfonts}
\usepackage{graphicx}

\begin{document}

\begin{center}

\large{\bf Coherent Interaction of Spins Induced by\\{}Thermal
Bosonic Environment}

\

\large{Dmitry~Solenov, Denis~Tolkunov and Vladimir~Privman}

\

Department of Physics, Clarkson University,\\{}Potsdam, New York
13699--5820, USA

\

\end{center}

\hrule

\begin{abstract}

We obtain and analyze the indirect exchange interaction between
two two-state systems, e.g., spins, in a formulation that also
incorporates the quantum noise that they experience, due to a
bosonic environment, for instance, phonons. We utilize a
perturbative approach to obtain a quantum evolution equation for
the two-spin dynamics. A non-perturbative approach is used to
study the onset of the induced interaction, which is calculated
exactly. We predict that for low enough temperatures the
interaction is coherent over time scales sufficient to create
entanglement, dominated by the zero-point quantum fluctuations of
the environment. We identify the time scales for which the spins
develop entanglement for various spatial separations.
\end{abstract}

\noindent{}Keywords: Exchange interaction; Entanglement; Thermal
environment; Qubit.

\

\noindent{}PACS: 75.30.Et, 71.70.$-$d, 71.55.$-$i, 73.21$-$b

\

\hrule

\

\

Studies of open quantum systems have a long history
\cite{Leggett,VKampen}. Recent experimental advances have allowed
the observation of fundamental quantum mechanical phenomena in
nanostructured systems in condensed matter and other fields.
Promise of applications for quantum information processing has
stimulated significant interest in theoretical studies of quantum
coherence and entanglement in situations when the quantum system
is subject to environmental noise
\cite{Leggett,VKampen,Tolkunov,Solenov,Agk}. In the present work,
we consider two two-state systems: qubits, e.g., spins
$\scriptstyle{1/2}$, in a thermal bosonic environment (bath). We
study the emergence of the indirect exchange interaction between
two localized spins and identify the regimes where entanglement
generated by this interaction can be observed. We report results
for geometries relevant for recent experiments
\cite{Xiao,Elzerman,Craig}.

Quantum computing schemes with qubits coupled by indirect exchange
of excitons were proposed in \cite{PVK,MPV,MPG,Piermarocchi}, and
these interactions are also of interest in studies of quantum
phase transitions \cite{Sachdev}. Traditionally, such RKKY-type
interactions were calculated perturbatively at zero temperature,
without considering quantum noise, with conduction electrons
\cite{RKKY} or exitons \cite{Bychkov} acting as the ``bath.''
Recent works \cite{Rikitake,Mozyrsky} considered the effects of
noise for a thermalized bath of, respectively, noninteracting and
interacting electrons. A bath of thermalized bosonic modes can
cause decoherence and, for more than one qubit, disentanglement.
These effects have been studied extensively in the recent
literature \cite{Dec2Qa}. It has also been anticipated
\cite{Dec2Qb} that such a thermalized bath of modes can induce
entanglement under certain conditions.

In this work, we investigate both physical effects of a
thermalized bosonic bath in which two qubits are immersed.
Specifically, the induced interaction, which is effectively a
zero-temperature effect, and the quantum noise, originating from
the same bath modes, are derived within a uniform treatment. We
study the dependence of  the induced interaction (coherent) vs.\
noise (decoherence) effects on the parameters of the bath modes,
qubits, and their coupling, as well as on the geometry of the
qubit system.

We consider two localized spins separated by distance $\mathbf{d}$
and identically coupled with the modes of a thermalized bosonic
bath, described by $H_B=\sum_{\mathbf{k}}\omega
_{\mathbf{k}}a_{\mathbf{ k}}^{\dagger}a_{\mathbf{k}}$, where we
set $\hbar=1$. The external magnetic field will be represented by
the Hamiltonian $H_S$, corresponding to the energy gap $\Delta$
between the up and down states for each spin. A natural example of
the described model are spins of two localized electrons
interacting via lattice vibrations (phonons) by means of the
spin-orbit interaction \cite{SO}. For each type of phonons, the
interaction can be assumed \cite{Leggett,VKampen,SO} linear in the
bosonic variables. Without loss of generality, the calculational
techniques can be developed for a coupling that involves a
specific spin component, which will be the same for both spins and
denoted by $S_j$ for the $j$th spin, localized at $\mathbf{r}_j$.
Thus we take $H_{SB}=\sum_{j=1,2}S_jX_j$, where
$X_j=\sum_{\mathbf{k}}g_{\mathbf{k}}e^{i\mathbf{k}\cdot\mathbf{r}_j}
\left(a_{\mathbf{k}}+a_{-\mathbf{k}}^{\dagger}\right)$. All our
derivations, however, can be generalized to include all the
projections of the spin \cite{STP:Big}, and we give some
illustrative numbers for such a calculation below. Our emphasis
here is on comparing the relative importance of the coherent vs.\
noise effects of a given bosonic bath in the two-qubit dynamics.
We do not include possible other two-qubit interactions in such
comparative calculation of dynamical quantities.

The overall system is described by the Hamiltonian
$H=H_S+H_B+H_{SB}$. Let us point out that such a model is quite
general and it also finds applications, for instance, in quantum
optics \cite{Scully} where the Hamiltonian $H$ would describe
atoms (regarded as the two level systems) interacting with an
electromagnetic field. Our detailed expressions here are obtained
for the one-dimensional case, relevant for recent experiments that
involve channel geometries \cite{Xiao,Elzerman,Craig}, when
phonons propagate along $\mathbf{d}$, i.e.,
$\mathbf{k}\cdot\mathbf{d} \to k |\mathbf{d}|$.

The problem of describing the dynamics of the spin system and,
thus, finding the reduced density matrix, $\rho_S(t)=Tr_B\rho(t)$,
cannot be solved exactly in the general case. We first consider
time scales longer that the thermalization times of the bath.
Furthermore, we treat the interaction, $H_{SB}$, as a
perturbation, and expand the equation for the density matrix,
$i\partial_t\rho(t)=\left[H,\rho(t)\right]$, up to the second
order in $H_{SB}$. In this regime, it is  appropriate
\cite{VKampen} to model the thermalization of the bath within a
standard Markovian approximation which involves factoring out the
environmental mode density matrix in the second order term,
replacing it by the thermal one, as well as using \cite{Leggett}
the initial condition $\rho (0)=\rho _S\left( 0\right) \otimes
\rho _B^{\rm thermal}$. Here $\rho _B^{\rm
thermal}=\prod_k\mathrm{Z}_k^{-1}e^{-\omega_k a_k^\dag a_k/k_BT}$,
and $\mathrm{Z}_k$ is the partition function for the oscillator
$k$. The resulting equation \cite{VKampen} for the density matrix
is
\begin{equation}\label{eq:SE}
i\partial _t\rho _S(t)=\left[ H_S,\rho _S(t)\right] +i\int_0^t
dt^{\prime }\Sigma \left(t^{\prime }-t\right) \rho _S(t),
\end{equation}
where $\Sigma \left( t^{\prime }-t\right) \rho _S\left( t\right)
=-\sum_{ij}Tr_B\left[ S_jX_j,\left[ S_i\left( t^{\prime }-t\right)
X_i\left( t^{\prime }-t\right) ,\rho _B^{\rm thermal}\rho _S\left(
t\right) \right] \right] $ is the self-energy superoperator term.

This expression involves the correlation functions
$C_{ji}(t)=Tr_B[ X_jX_i(t)$ $\times\rho _B^{\rm thermal}]$, where
$i,j = 1,2$ for the two spins, with the property
$C_{ji}^{*}(t)=C_{ji}(-t)$. The present approximation assumes
\cite{VKampen} that the bath has very short memory. One can argue
that the correlation functions $C_{ji}(t)$ are nonnegligible only
up to times $1/\omega_c$, where $\omega_c$ is the frequency cutoff
for electron spins interacting with phonons. This cutoff comes
either from the phonon density of states or from the localization
of the electron wavefunctions. This time scale will be considered
to be significantly smaller then the system evolution times
defined by the inverse of the energy gap $\Delta$.

In the resulting evolution equation,
\begin{equation}\label{eq:MA}
i\partial _t\rho _S(t)=\left[ H_{\rm eff},\rho _S(t)\right]
+i\mathcal{L}_c\rho _S(t)+i\mathcal{L}_s\rho _S(t),
\end{equation}
we separate out the coherent dynamics in the first term. The
superoperators $\mathcal{L}_c$ and $\mathcal{L}_s$ will be
addressed shortly. In the two-spin case, one can establish by a
lengthy calculation \cite{STP:Big}, which
is not reproduced here, that
\begin{equation}\label{eq:Heff}
H_{\rm eff}=H_S+2\chi _c(\mathbf{d})S_1S_2+O\left( \chi
_s(\mathbf{d}),\eta _s(0)\right),
\end{equation}
which includes the effective coupling of the spin components.
Expressions for the quantities $\chi _{c,s}$ and $\eta _{c,s}$
follow from, respectively, the imaginary and real parts of the
correlation functions and will be given explicitly below. The last
term in (\ref{eq:Heff}) introduces corrections of relative order
$\Delta / \omega_c$ in the induced interaction and is not of
interest here.

The second term in (\ref{eq:MA}) is
\begin{equation}\label{eq:LC}
\mathcal{L}_c\rho _S(t)=-\sum_{ij}\eta _c\left( \theta
_{ij}\mathbf{d} \right) \left\{ S_iS_j,\rho _S(t)\right\}
+2\sum_{i\neq j}\eta _c\left( \theta _{ij}\mathbf{d}\right)
S_i\rho _S(t)S_j,
\end{equation}
where $\theta _{ij}\equiv 1-\delta _{ij}$; it accounts for the
dominant relaxation and decoherence processes. The third term in
(\ref{eq:MA}), $\mathcal{L}_s\rho _S(t)$, involves expressions
proportional to $\eta _s\left( \theta _{ij}\mathbf{d}\right) $ or
$ \chi _s\left( \theta _{ij}\mathbf{d}\right)$ and may often be
considered small compared to terms in (\ref{eq:LC}), as will be
shown bellow by analyzing the magnitudes of $\eta _s\left( \theta
_{ij}\mathbf{d}\right) $ and $ \chi _s\left( \theta
_{ij}\mathbf{d}\right)$.

In (\ref{eq:Heff}) and (\ref{eq:LC}) we introduced the quantities
\begin{equation}\label{eq:xc}
\chi _c\left( \mathbf{d}\right) =\int\limits_{-\infty }^0 {\!}dt\int\limits_0^\infty
{\!}d\omega \, \mathcal{D}\left( \omega \right) \left| g\left( \omega
\right) \right| ^2\sin \omega t\cos \frac{\omega|\mathbf{d}|}{c_s}\cos
\Delta t
\end{equation}
and
\begin{equation}\label{eq:etc}
\eta _c\left( \mathbf{d}\right) = \int\limits_{-\infty
}^0{\!}dt\int\limits_0^\infty {\!}d\omega \, \mathcal{D}\left(
\omega \right) \left| g\left( \omega \right) \right| ^2\coth \frac
\omega {2k_BT}\cos \omega t\cos \frac{\omega|\mathbf{d}|}{c_s}\cos
\Delta t,
\end{equation}
where $\mathcal{D}\left( \omega \right) $ is the density of states
of the bosonic modes; this factor should also include the Debye
cutoff at large frequencies. For definiteness, we have assumed the
linear dispersion, $\omega =c_sk$, because details of the
dispersion relation for larger frequencies usually have little
effect on decoherence properties. Another reason to focus on the
low-frequency modes is that an additional cutoff, $\omega_c$,
resulting from the localization of the electron wave functions,
typically much smaller than the Debye frequency, will be present
due to the factors $\left| g\left( \omega \right) \right| ^2$. The
expressions for $\eta _s\left( \mathbf{d}\right) $ and $\chi
_s\left( \mathbf{d}\right) $ can be obtained by replacing $\cos
\Delta t\rightarrow \sin \Delta t$ in $\eta _c\left(
\mathbf{d}\right) $ and $\chi _c\left( \mathbf{d}\right) $,
respectively. The integration in (\ref{eq:etc}) then yields
\begin{equation}\label{eq:etcF}
\eta _c\left( \mathbf{d}\right) =\frac \pi 2\mathcal{D}\left(
\Delta \right) \left| g\left( \Delta \right) \right| ^2\coth \frac
\Delta {2k_BT}\cos \frac{\Delta |\mathbf{d}|}{c_s}.
\end{equation}
Similarly, we find
\begin{equation}\label{eq:xs}
\chi _s\left( \mathbf{d}\right) =\frac \pi 2\mathcal{D}\left(
\Delta \right) \left| g\left( \Delta \right) \right| ^2\cos
\frac{\Delta |\mathbf{d}|}{c_s}.
\end{equation}

To derive explicit expressions for $\chi _c\left( \mathbf{d}
\right) $, one needs to specify the $\omega$-dependence in
$\mathcal{D}\left( \omega \right) \left| g\left( \omega \right)
\right| ^2$. For purposes of modeling bosonic heat-bath effects,
this product is usually approximated \cite{Leggett} by a power law
with superimposed exponential cutoff, $\mathcal{D}\left( \omega
\right) \left| g\left( \omega \right) \right| ^2=\alpha _n\omega
^ne^{-\omega /\omega _c}$, with $n\geq 1$. Finally, we get
\begin{equation}\label{Deriv}
\chi _c\left( \mathbf{d}\right) =-\alpha _n\omega _c^n\left[ \xi
^n\frac{\partial ^{n-1}}{\partial \xi ^{n-1}}\frac{\left(
-1\right) ^{n-1}\xi }{ 1+\xi ^2}\right] _{\xi =c_s/\left( \omega
_cd\right) } \, .
\end{equation}
This quantity gives the coefficient of the leading induced
interaction; see (\ref{eq:Heff}). The individual Lamb shifts in
(\ref{eq:Heff}) are defined by $\eta _s(0)$ and are not important
for our discussion.

For the commonly studied case of Ohmic dissipation \cite{Leggett},
$n=1$, the environment induces the following spin-spin
interaction,
\begin{equation}\label{HintM}
H_{\rm int}=-\frac{2\alpha _1\omega _c}{1+\omega
_c^2\mathbf{d}^2/c_s^2}S_1S_2\,{}.
\end{equation}
This induced interaction is temperature independent and is
mediated by the zero-point fluctuations of the bosonic field. It
is long-range and decays as a power-law for large $|\mathbf{d}|$.
On the other hand, quantum noise terms, see
(\ref{eq:MA}-\ref{eq:LC}), depend weakly on $|\mathbf{d}|$, and
increase with temperature. In Fig.~1, we plot the magnitude of the
interaction $\chi _c\left( \mathbf{d}\right) $ as a function of
the spin separation. It is compared, for varying temperature, to
the magnitude of the decoherence terms, $\chi _s\left(
\mathbf{d}\right) $, $\eta _s\left( \mathbf{d}\right)$ and $\eta
_c\left( \mathbf{d}\right) $, among which $\eta _c\left(
\mathbf{d} \right) $ is the dominant for the Ohmic case, $n=1$,
and the temperature scale of Fig.~1, i.e. for $2kT/\omega_c$ from
$\sim 0.05$. For lower temperatures evaluation of $\eta _s\left(
\mathbf{d}\right)$ may be important. For the super-Ohmic case,
$n>1$, one obtains similar behavior, except that the interaction
decays as a higher negative power of $|\mathbf{d}|$.

\begin{figure}
\begin{center}
\includegraphics[width=8cm]{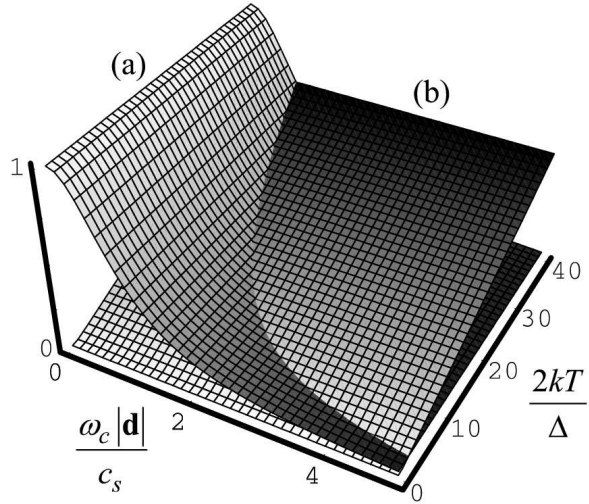}
\caption{The magnitude, in units of $\alpha _1\omega _c$, of (a)
the induced spin-spin interaction, and (b) the largest decoherence
amplitude. Here we took $\Delta/\omega_c=0.01$.}
\end{center}
\end{figure}

The amplitudes plotted in Fig.~1 provide qualitative information
on the dynamics. For definiteness, let us consider the case of
$H_S={\scriptstyle{1\over 2}}\Delta \sigma _z^{\left( 1\right)
}+{\scriptstyle{1\over 2}}\Delta \sigma _z^{\left( 2\right) }$ and
$H_{\rm int}$ in (\ref{HintM}) involving $S_{1,2} = \sigma
_x^{\left( 1,2\right)}$. If the induced interaction were the only
effect of the bath, then the system's dynamics would be coherent,
with oscillations determined by the energy gaps of $H_S+H_{\rm
int}$. This Hamiltonian has the singlet state $\left( \left|
\uparrow \downarrow \right\rangle -\left| \downarrow \uparrow
\right\rangle \right) /\sqrt{2} $, with the energy $E_2=-2\chi
_c\left( \mathbf{d}\right) $, and the (split) ``triplet''
$C\left(\left| \uparrow \uparrow \right\rangle - \delta \left|
\downarrow \downarrow \right\rangle\right)$, $\left( \left|
\uparrow \downarrow \right\rangle +\left| \downarrow \uparrow
\right\rangle \right)/\sqrt{2}$, and $C\left(\left| \downarrow
\downarrow \right\rangle + \delta \left| \uparrow \uparrow
\right\rangle\right)$, with the energies $E_0=-\sqrt{\Delta^2
+4\chi^2_c\left( \mathbf{d}\right)}$, $E_1=2\chi _c\left(
\mathbf{d}\right) $, and $E_3=\sqrt{\Delta^2 +4\chi^2_c\left(
\mathbf{d}\right)}$, respectively, where $\delta \equiv 2\chi
_c\left( \mathbf{d}\right)/\left(\Delta-E_0\right)$, and $C$ is
the normalization constant. The effect of the noise terms is to
wash away the coherent behavior. The time scales of the coherent oscillations
and of the noise-induced relaxation processes, defined by the corresponding frequencies in Fig.~1,
become comparable at the intersection curve of the two surfaces in
Fig.~1. For larger distances and/or temperatures the noise terms dominate.

The energy gap between $E_1$ and $E_2$ is defined by $\chi
_c\left( \mathbf{d}\right) $, whereas the effective width of each
level due to decoherence will be determined by the magnitudes of
$\chi _s\left( \mathbf{d}\right) $, $\eta _s\left(
\mathbf{d}\right) $ and $\eta _c\left( \mathbf{d}\right) $. Let us
estimate the magnitude of this energy splitting. As an example, we
consider two phosphorus donor impurities in Ge, in external
magnetic field in the $z$ direction. Due to the symmetry of the
spin-phonon coupling (via spin-orbit interaction), only two
components, namely, $\sigma_x$ and $\sigma_y$, are important
\cite{MKGB}. One can demonstrate that in this case the terms in
(\ref{eq:SE}) proportional to products of $\sigma_x$ and
$\sigma_y$ vanish, while the contributions from the $\sigma_x$ and
$\sigma_y$ terms are identical. Therefore, we can use the
single-component results, with $S_j\sim\sigma^{(j)}_x$. For a
donor impurity electron spin, the cutoff $\omega_c\sim c_s/a_B$
comes from the donor electron wave function \cite{MKGB}, which is
localized on the scale of $a_B\sim 4$nm. The characteristic value
of the phonon group velocity in Ge is about $3\times 10^3\,$m/s.
The strength of the spin-orbit coupling, $\alpha_1 \omega_c$, is
\cite{MKGB} of the order of $10^7$s$^{-1} $. Utilizing these data
in (\ref{eq:etcF},\ref{HintM}), with $n=1$ which, strictly
speaking should be only valid for one-dimensional channel for
phonon propagation, we obtain the estimate for the
$E_1\leftrightarrow E_2$ energy splitting of about $10\,$MHz,
whereas the noise level for mK temperatures varies from $0.1$ to
$1\,$MHz, depending on the magnitude of the Zeeman splitting. This
coherence/noise ``measure'' can be further improved for different
cases of the phonon spectrum, the shape of the wave function,
etc.\ \cite{STP:Big}.

To study the onset of the exchange interaction, we consider the
case when the Zeeman splitting $\Delta$ is negligible. Then one
can actually derive an exact solution for $H_S + H_B + H_{SB}$,
with the bath modes traced over without the Markovian assumption,
and demonstrate the emergence of the effective interaction
Hamiltonian as in (\ref{eq:Heff},\ref{Deriv},\ref{HintM}). For
$\Delta = 0$, a lengthy calculation utilizating bosonic operator
techniques yields
\begin{equation}\label{eq:dme}
\rho _S\left( t\right) =\sum_{\mathbf{\lambda },\mathbf{\lambda
}^{\prime}}P_{\mathbf{\lambda }}\rho _S\left( 0\right)
P_{\mathbf{\lambda }^{\prime}}e^{\mathcal{L}_{\mathbf{\lambda
\lambda }^{\prime }}\left( t\right) },
\end{equation}
where $\left|\lambda _j\right\rangle$ is an eigenstate of $S_j$
labeled by its eigenvalue $\lambda _j$, and we introduced the
projection operator $P_{\mathbf{\lambda }}=\left|\lambda _1\lambda
_2\right\rangle \left\langle \lambda _1\lambda _2\right|$. The
exponent consists of the real part, which represents decoherence,
\begin{figure}
\begin{center}
\includegraphics[width=8cm]{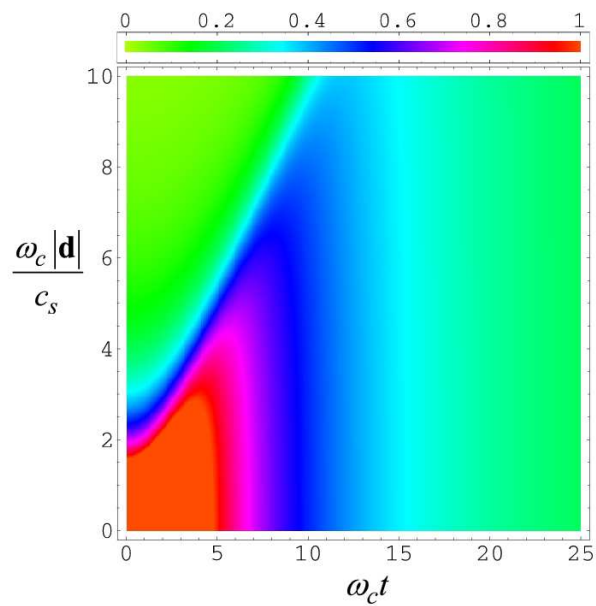}
\caption{The onset of the indirect spin-spin interaction is
measured by the decay of the correction term $F(t)$. The values of
$F(t)$ are in units of $\alpha_1 \omega_c$, and are color-coded
according to the top bar.}
\end{center}
\end{figure}
%
%
\begin{eqnarray}\label{eq:res}\nonumber
\mathrm{Re}\mathcal{L}_{\mathbf{\lambda \lambda }^{\prime }}\left(
t\right) = &-& \sum_kG_k(t;T)\left[ \left( \lambda _1^{\prime
}-\lambda _1\right) ^2+\left( \lambda _2^{\prime }-\lambda
_2\right) ^2\right. \\&+& \Bigl. 2\cos (\omega
_k|\mathbf{d}|/c_s)\left( \lambda _1^{\prime }-\lambda _1\right)
\left( \lambda _2^{\prime }-\lambda _2\right)\Bigr]\, ,
\end{eqnarray}
and imaginary part, which describes the coherent evolution,
\begin{equation}\label{eq:ims}
\mathrm{Im}\mathcal{L}_{\mathbf{\lambda \lambda }^{\prime }}\left(
t\right)=\sum_k C_k(t)\cos \left( \omega _k|\mathbf{d}|/c_s\right)
\left( \lambda _1\lambda _2-\lambda _1^{\prime }\lambda _2^{\prime
}\right)\, .
\end{equation}
Since for $\Delta =0$, $H_S$ commutes with $H_{SB}$, quantum
noise in this case only affects the off-diagonal matrix elements. Eventually,
this destroys quantum correlations between the spins. In
(\ref{eq:res}) and (\ref{eq:ims}) the standard functions
\cite{Leggett,VKampen,Tolkunov} were introduced,
\begin{equation}\label{eq:gk}
G_k(t;T)=2\frac{\left| g_k\right| ^2}{\omega _k^2}\sin
^2\frac{\omega _kt} 2\coth \left(\omega _k/2k_BT\right),
\end{equation}
\begin{equation}\label{eq:ck}
C_k(t)=2\frac{\left| g_k\right| ^2}{\omega _k^2}\left( \omega
_kt-\sin\omega _kt\right) .
\end{equation}

Our focus here is on the imaginary part (\ref{eq:ims}). One can
demonstrate that if the noise terms (the real part) were
completely absent, the resulting evolution would be coherent with
the evolution operator given by $e^{-i\left[H_{\rm int}+F(t)
\right]t}$, where
\begin{equation}\label{eqF}
F(t)=2S_1S_2\int_0^\infty \!d\omega \frac{\mathcal{D}\left( \omega
\right) \left| g(\omega )\right| ^2}\omega \frac{\sin \omega
t}{\omega t} \cos(\omega |\mathbf{d}|/c_s).
\end{equation}
and
\begin{equation}\label{HintA}
H_{\rm{int}}=-\frac{2\alpha_n \Gamma \left( n\right) \omega
_c^n}{\left( 1+\omega _c^2\mathbf{d}^2/c_s^2\right) ^{n/2}}\cos
\left[ n\arctan \left( \frac{\omega_c|\mathbf{d}|}{c_s}\right)
\right]S_1S_2 .
\end{equation}
Expression (\ref{eqF}) represents the initial, time-dependent
correction present only during the onset of the induced
interaction. Specifically, $F(0)=-H_{\rm int}$, but $F(t)$ decays
for large times. In Fig.~2, we plot $F(t)$  for the Ohmic case,
$n=1$, as a function of time and spin separation. The right side
of the plot in Fig.~2 corresponds to the constant coupling regime
with the interaction Hamiltonian (\ref{HintA}). Note that the
Hamiltonian (\ref{HintA}) with $n\ge1$ is identical to that
obtained within the perturbative Markovian approach, cf.
(\ref{eq:Heff},\ref{Deriv},\ref{HintM}).

\begin{figure}\label{fig3}
\begin{center}
\includegraphics[width=8cm]{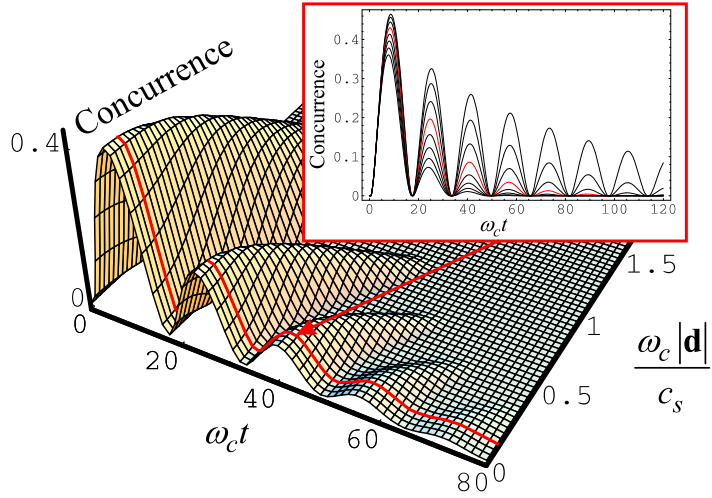}
\caption{The concurrence calculated for Ohmic dissipation with
$\alpha_1=k_B T/\omega_c=1/20$. The inset gives the time
dependence for different temperatures: $80 k_B T/\omega_c = 1, 2,
3, 4, 5, 6, 7, 8$, from the top curve to the bottom one,
respectively: the top curve corresponds to the lowest
temperature.}
\end{center}
\end{figure}

Let us investigate the role of the decoherence resulting from the
real part of $\mathcal{L}_{\mathbf{\lambda \lambda }^{\prime }}$
in (\ref{eq:res}). Note that within the exact solution the bath is
thermalized only initially, at $t=0$. However, it is expected that
the effects of the quantum noise are represented {\it
qualitatively\/} similarly to the Markovian approximation, see
(\ref{eq:MA}-\ref{eq:LC}), which    implies re-thermalization of
the bath after each infinitesimal time step. To analyze the
dynamics of quantum correlations between the spins, we have used the
concurrence \cite{Wootters} --- a measure of entanglement which
is widely used in quantum information theory. For a mixed
state of two qubits it is defined \cite{Wootters} via the
eigenvalues $\lambda_j$ of $\sqrt{\sqrt{\rho_S}\sigma^1_y
\sigma^2_y \, \rho _S^* \, \sigma^1_y \sigma^2_y\sqrt{\rho_S}}$ as
$\max\{\,0,2\mathop {\max }\limits_i
\lambda_i-\lambda_1-\lambda_2-\lambda_3-\lambda_4\,\}$. For
$\Delta =0$, and as long as the effects of the quantum noise are
small, the induced interaction will split the system energies into
two degenerate pairs, $E_0 = E_2$ and $E_1 = E_3$. As a result,
the dynamics will involve coherent oscillations with the frequency
defined by the gap $E_1 - E_0 = 4 \chi _c\left(
\mathbf{d}\right)$. In Fig.~3, we plot the concurrence for the
density matrix given by (\ref{eq:dme}), as a function of time and
spin-spin separation, for the (initially unentangled) state
$\left|\uparrow\uparrow\right\rangle$, and $n=1$. The inset in
Fig.~3 shows the time dependence of the concurrence for different
temperatures. Figure~3 demonstrates that the system can develop
and maintain entanglement over several coherent-dynamics
oscillations before the noise-induced effects take over and the
concurrence decays to zero.

In summary, we studied the induced indirect exchange interaction
due to a bosonic bath of environmental modes which also introduce
the noise. We demonstrated that it can create {\it observable\/}
two-spin entanglement. For an appropriate choice of the system
parameters, specifically, the spin-spin separation, this
entanglement can be maintained and the system can evolve
approximately coherently for many cycles of its internal dynamics.
However, for large times the quantum noise effects will eventually
dominate and the entanglement will be erased.

The authors acknowledge useful discussions with
and helpful suggestions by J.~Eberly, L.~Fedichkin and D.~Mozyrsky. This research was supported by the NSF under grant
DMR-0121146.


\begin{thebibliography}{99}{\frenchspacing

\bibitem{Leggett} A. J. Leggett, S. Chakravarty, A. T. Dorsey, M. P. A. Fisher, A.
Garg, and W. Zwerger, Rev. Mod. Phys. \textbf{59}, 1 (1987).

\bibitem{VKampen} N. G. van Kampen, \emph{Stochastic Processes in Physics and Chemistry\/}, (North-Holland, Amsterdam, 2001).

\bibitem{Tolkunov}  D. Tolkunov and V. Privman, Phys. Rev. A \textbf{69}, 062309 (2004); D. Tolkunov and V. Privman, Phys. Rev. A \textbf{71}, 060308 (2005).

\bibitem{Solenov}  D. Solenov and V. Privman,  Int. J. Modern Phys. B \textbf{20}, 1476 (2006).

\bibitem{Agk} D. Gobert, J. Delft, and V. Ambegaokar, Phys. Rev. A \textbf{70}, 026101 (2004).

\bibitem{Xiao} M. Xiao, I. Martin, E. Yablonovitch, and H. W. Jiang, Nature \textbf{430}, 435 (2004).

\bibitem{Elzerman} J. M. Elzerman, R. Hanson, L. H. Willems van Beveren, B. Witkamp, L. M. K. Vandersypen, and L. P. Kouwenhoven, Nature \textbf{430}, 431 (2004).

\bibitem{Craig} N. J. Craig, J. M. Taylor, E. A. Lester, C. M. Marcus, M. P. Hanson, and A. C. Gossard, Science \textbf{304}, 565 (2004).

\bibitem{PVK} V. Privman, I. D. Vagner, and G. Kventsel, Phys. Lett. A \textbf{239}, 141 (1998).

\bibitem{MPV} D. Mozyrsky, V. Privman, and I. D. Vagner, Phys. Rev. B \textbf{63}, 085313 (2001).

\bibitem{MPG} D. Mozyrsky, V. Privman, and M. L. Glasser, Phys. Rev. Lett. \textbf{86}, 5112 (2001).

\bibitem{Piermarocchi} C. Piermarocchi, P. Chen, L. J. Sham, and D. G. Steel, Phys. Rev. Lett. \textbf{89}, 167402 (2002).

\bibitem{Sachdev} S. Sachdev, \emph{Quantum Phase Transitions\/} (Cambridge University Press, 1999).

\bibitem{RKKY}  M.A. Ruderman, C. Kittel, Phys. Rev. \textbf{96}, 99 (1954); T. Kasuya, Prog. Theor. Phys. \textbf{16}, 45 (1956);
K. Yosida, Phys. Rev. \textbf{106}, 893 (1957).

\bibitem{Bychkov} Yu. A. Bychkov, T. Maniv and I. D. Vagner, Solid State Comm. \textbf{94}, 61 (1995).

\bibitem{Rikitake} Y. Rikitake, H. Imamura, Phys. Rev. B \textbf{72}, 033308 (2005).

\bibitem{Mozyrsky} D. Mozyrsky, A. Dementsov, and V. Privman, Phys. Rev. B \textbf{72}, 233103 (2005).

\bibitem{Dec2Qa} T. Yu and J. H. Eberly, Phys. Rev. Lett.
\textbf{93}, 140404 (2004); T. Yu and J. H. Eberly, Phys. Rev. B
\textbf{68}, 165322 (2003).

\bibitem{Dec2Qb} see, for example, D. Braun, Phys. Rev. Lett. \textbf{89}, 277901
(2002); D. Porras and J. I. Cirac, Phys. Rev. Lett. \textbf{92},
207901 (2004).

\bibitem{SO} G. D. Mahan, \emph{Many-Particle Physics\/} (Kluwer Academic, 2000);
R. Winkler, \emph{Spin-Orbit Coupling Effects in Two-Dimentional
Electron and Hole Systems\/} (Springer, 2003).

\bibitem{STP:Big} D. Solenov, D. Tolkunov, and V. Privman, e-print:  cond-mat/0605278.

\bibitem{Scully} M. O. Scully and M. S. Zubairy \emph{Quantum Optics\/} (Cambridge University Press, 1997).

\bibitem{MKGB} D. Mozyrsky, Sh. Kogan, V. N. Gorshkov, and G. B. Berman, Phys. Rev. B \textbf{65}, 245213 (2002).

\bibitem{Wootters}  S. Hill and W. K. Wootters, Phys. Rev. Lett. \textbf{78}, 5022 (1997);
W. K. Wootters, Phys. Rev. Lett. \textbf{80}, 2245 (1998).

}\end{thebibliography}
\end{document}